\documentclass[psfig,useAMS,usenatbib]{mn2e}
\usepackage{aas_macros}
\usepackage{graphicx}
\usepackage{epstopdf}

\title{Optimal Surveys for Weak Lensing Tomography}
\author[A. Amara \& A. R{\'e}fr{\'e}gier]
{Adam Amara \& Alexandre R{\'e}fr{\'e}gier\\
Service d'Astrophysique, CEA Saclay, 91191 Gif sur Yvette, France. (email: amara@cea.fr)}
\date{\today}
\begin{document}

\maketitle

\label{firstpage}

\begin{abstract}
Weak lensing surveys provide a powerful probe of dark energy through the
measurement of the mass distribution of the local Universe. A number of ground-based
and space-based surveys are being planned for this purpose. Here, we study the optimal 
strategy for these future surveys using the joint constraints on the equation of state parameter $w_n$ and its evolution $w_a$ as a figure of merit by considering power spectrum tomography. For this purpose, we first consider an `ideal' survey which is both wide and deep and exempt from systematics. We find that such a survey has great potential for dark energy studies, reaching one sigma precisions of $1\%$ and $10\%$ on the two parameters respectively. We then study the relative impact of various limitations by degrading this ideal survey. In particular,  we consider the effect of sky coverage, survey depth, shape measurements systematics, photometric redshifts systematics and uncertainties in the non-linear power spectrum predictions. 
We find that, for a given observing time, it is always advantageous to choose a wide rather than a deep survey geometry. We also find that the dark energy constraints from power spectrum tomography
are robust to photometric redshift errors and catastrophic failures, if a spectroscopic calibration sample of $10^4-10^5$ galaxies is available. The impact of these systematics is small compared to the limitations that come from potential uncertainties in the power spectrum, due to shear measurement and theoretical errors.  To help the planning of future surveys, we summarize our results with comprehensive scaling relations which avoid the need for full Fisher matrix calculations.

\end{abstract}


\section{Introduction}
Over the past decade, cosmic shear measurements have proved to be powerful tools for probing the properties of the observable Universe.  Current cosmic shear measurements typically cover areas of a few to a few tens of square degrees \citep{2004astro.ph..4195M,2004MNRAS.347..895H,2004ApJ...605...29R,2006ApJ...647..116H,2006A&A...452...51S} and are able to measure the amplitude of the matter power-spectrum to an accuracy of $\sim 10\%$.  For these surveys, the dominant uncertainties come from the statistical errors (cosmic variance)arising from the limited survey area coverage and observational  systematics, such as uncertainties in the Point Spread Function (PSF). Within the next decade, a number of surveys have been proposed (such as DUNE\footnote{http://www.dune-mission.net}, PanSTARRS\footnote{http://pan-starrs.ifa.hawaii.edu}, DES\footnote{https://www.darkenergysurvey.org}, VISTA\footnote{http://www.vista.ac.uk}, SNAP\footnote{http://snap.lbl.gov} and LSST\footnote{http://www.lsst.org}) that will observe several thousand square degrees with the aim of constraining the matter power-spectrum with much greater accuracy.  Survey areas of this size overcome many of the problems of statistical uncertainties, so the immediate challenge facing these surveys is to develop techniques for controlling systematic errors.

One of the major goals of future large area cosmic shear surveys is to understand the nature of dark energy by placing tight constraints on the equations of state parameter, w, which is the ratio of pressure to the density of the dark energy. These future lensing surveys should be able to place constraints on both the overall value of the equation of state and also its evolution with time by using the redshift information of the lensed galaxies. However, understanding the impact of systematic errors is still in need of a great deal of study.  In this paper, we investigate how future weak lensing surveys can be optimised to yield the strongest constraints on dark energy. For this purpose, we make use of the figure of merit (FOM) proposed by the Dark Energy Task Force (DETF)\footnote{http://www.nsf.gov/mps/ast/detf.jsp} to compare the strength of future dark energy surveys with different cosmological probes. For our optimisation study, we will only consider weak lensing power spectrum tomography, while other weak lensing statistics, such as the bispectrum and cross-correlation cosmography, have been shown to provide additional cosmological constraints \citep{2003PhRvL..91n1302J,2004ApJ...600...17B,2006astro.ph..6416T}. Although these other techniques could provide extra information and may require a different optimisation, we choose lensing tomography for optimisation since it is the most tested and robust technique and as a single method offers the strongest constraints. 

We thus investigate the impact of three distinct potential limitations of future surveys: (i) the statistical error imposed by a specific survey geometry; (ii) photo-z errors; and (iii) uncertainties in the power-spectrum arising from shear measurement systematics and theoretical uncertainties.   We  assume that cosmic variance errors are Gaussian, which does not hold on small scales.   The impact of full non-Gaussian errors still needs to be quantified further.

This paper is organised as follows. In section 2, we set out the background theory for our dark energy models, weak lensing tomography and Fisher matrix formalism.  In section 3, we set out the properties of our benchmark surveys.  The impact of limitations, such as finite observing time and systematics errors, are explored in section 4 and discussed in section 5. We conclude and summarise our work in section 6. Finally, appendix A describes our model for incorporating photometric redshift errors.

\section{Theory}

\subsection{Dark Energy model}
\label{sec:theory_DEM}
The impact of dark energy on gravitational lensing has been widely studied \citep{1999ApJ...514L..65H,2002PhRvD..65f3001H,2003PhRvL..91n1302J,2004PhRvD..70f3510S,2005MNRAS.363..469I,2006ApJ...636...21M,2006astro.ph..6416T}, with the effect depending on the exact nature of the dark energy considered. In this work, we consider the usual parametrisation \citep{2001IJMPD..10..213C,2003PhRvL..90i1301L} of evolution of the dark energy equation of state parameter, the $w = p/\rho$, i.e.
\begin{equation}
\label{ }
w(a)=w_n +(a_n-a)w_a,
\end{equation}
where $a=(1+z)^{-1}$ is the scale factor and $a_n$ the point at which the taylor expansion of $w(a)$ is performed \citep{2005ASPC..339..215H}.   

Since we wish to compare a variety of methods and configurations, a standard measure of the quality of a survey must be agreed upon. To this end, the FOM has emerged as a useful standard. The Figure is proportional to the inverse of the area of the $2\sigma$ ellipses in the $w_a-w_n$ plane. This quantity can be calculated either by setting $a_n$ to be the `pivot' point (i.e. the points where $w_n$ and $w_a$ are uncorrelated), in which case FOM $= 1/(4\Delta w_n \Delta w_a$).  A more straightforward method is to calculate the FOM using the square root of inverse of the determinant of the covariance matrix containing the dark energy parameter ($w_a$ and $w_n$). 

Our decision to use the FOM as a measure of quality of future survey is driven by the fact that we believe future surveys should be judged on their ability to reduce the error bars on a two paramter w model.  Other possible measures of quality exist.  \cite{2006astro.ph..8269A}  show that future lensing surveys should have small errors, i.e. smaller than unity, for a w model with 9 eigen values.  Since any one of these eigen values could be used as a diagnostic tool to test $\Lambda$CDM, one could argue that all 9 elements should be included in the measure of quality of future surveys.  This 9D-FOM could be expected to scale as the $9D-FOM \propto FOM^{9/2}$.   An alternative approach is to optimise directly using the errors on each w parameter.  For instance, one could optimise using the errors on a constant w model (as show in figure \ref{fig:deep_v_wide}).  A figure of merit constructed from the errors on one w parameter, 1D-FOM, would be expected to scale as the root of the FOM used in this paper ($1D-FOM \propto FOM^{1/2} $).  It should be noted that the FOM itself is not an error estimate for a given parameter. Instead it is an attempt to form a metric that quantifies the 'goodness' of a survey.  As we have illustrated this depends on both the number of parameters that a survey is able to constrain (i.e. errors smaller than unity) and the errors on those parameters.  Clearly if one is interested in the errors on an individual dark energy paramter one would need to marginalise over all other parameters.  These errors would then tend to scale in a similar way to the 1D-FOM. The exact choice of FOM will only enhance the relative merit of one survey over another and will not change which one is prefered.  The DETF report uses a two parameter figure of merit, which is becoming the reference in the community. We, therefore, have chosen to stick consistently to the 2 parameter FOM described above.

\subsection{Weak lensing tomography}
To evaluate the accuracy of weak lensing surveys, we use the power spectrum
tomography formalism described by \cite{2004PhRvD..70d3009H}. In this formalism, the background lensed galaxies are divided into redshift slices. The power spectrum corresponding to the correlations of shears, both within the slices and between slices, is then used to constrain cosmology models. We ignore the additional information provided by galaxy counts, such as the galaxy-shear correlation function. Our cosmological models are calculated using a BBKS \citep{1986ApJ...304...15B} transfer function to calculate the linear power spectrum and \citet{1996MNRAS.280L..19P} for the non-linear correction. The 3D power spectrum is projected into a 2D lensing correlation function using Limbers equation.

\subsection{Fisher Matrix}
The Fisher matrix for the shear power spectrum tomography is given by \citep{2004PhRvD..70d3009H}
\begin{equation}
\label{ }
F_{\alpha\beta}=f_{\rm sky}\sum_l\frac{(2l+1)\Delta l}{2}{\rm Tr}[D_{l\alpha}\widetilde{C}^{-1}_lD_{l\beta}\widetilde{C}^{-1}_l], 
\end{equation}
where the sum is over bands of multipole $l$ of width $\Delta l$, ${\rm Tr}$ 
stands for the trace, and
$f_{\rm sky}$ is the fraction of sky covered by the survey. The observed shear power
spectra,
\begin{equation}
\label{ }
\widetilde{C}_l^{ij} = C_l^{ij} +N_l^{ij}, 
\end{equation}
for each pair $i,j$ of redshift bins are written as a sum of the lensing and noise power spectra, respectively. The derivative matrices are given by
\begin{equation}
\label{ }
[D_{l\alpha}]^{ij} = \frac{\partial C_l^{ij}}{\partial p_\alpha},
\end{equation}
where $p_{\alpha}$ is the vector of the model parameters considered.

In this study, we use a 7 parameter fiducial flat cosmological model $p_{\alpha}=[\Omega_m = 0.28,  w_n = -0.95, w_a=0.00, h=0.72, \sigma_8 = 1.0, \Omega_b = 0.046, n=1.0]$. When we determine the errors we expect on the $\rm w_n$ and $\rm w_a$, we marginalise over the other 5 parameters without any external priors.

\section{Benchmark Surveys}
\label{sec:benchmark}
For our optimisation analysis, we consider several benchmark surveys which we will use as references to study the impact of potential limiting factors and systematics. The first benchmark survey, which we call the `ideal' survey that is both wide and deep.  This is chosen to be ultra-wide, i.e. to cover the half of the sky that is not obscured by the Milky Way corresponding to a coverage of 20,000 square degrees like DUNE.  The depth of this ideal survey correspond to the depth of the SNAP wide survey \citep{2004AJ....127.3089M,2004AJ....127.3102R}.  Such a survey can be expected to have 100 galaxies per amin$^2$ with a median redshift of $\rm z_m=1.23$. The survey is assumed to have no shape measurement systematics errors and no error imposed on the predictions from theory and no photometric redshift errors. This survey can be considered to be ideal in the sense that, using a space-based like instrument, such as SNAP, it would take roughly thirty years to collect this data.  For our benchmark calculations, we also use the the scales in the range $10<\ell<2\times10^4$. We calculate the FOM for this survey, using cosmic shear tomography by dividing the galaxies into 10 redshift bins, and find $FOM=220$.  

\begin{figure} 
\centering
\resizebox{0.95\columnwidth}{0.95\columnwidth}{\includegraphics{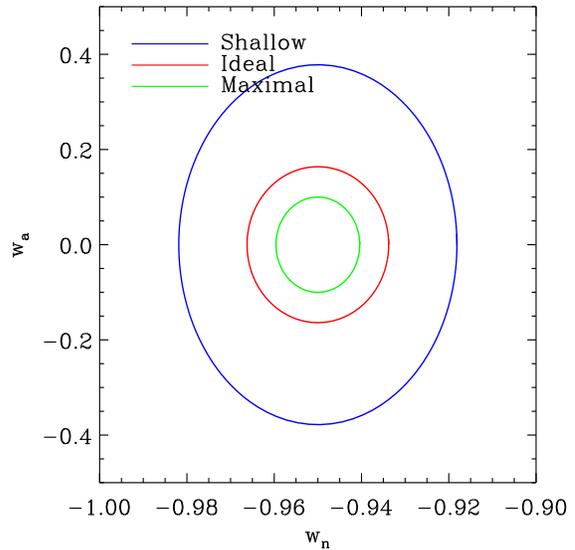}}
\caption{Dark energy constraints for a number of possible ideal surveys. The ideal survey covers half the sky (20,000 sq. degrees) with a median redshift of $z_m=1.23$ and 100 galaxies per amin$^2$.  Also shown is a maximal ideal survey with  $z_m=1.43$ and 260 galaxies per amin$^2$.  Finally, a shallow survey is shown that has 35 galaxies per amin$^2$  and $z_m=0.9$}

\label{fig:ideal} 
\end{figure}

For comparison, we explore the effect of deviations from this `ideal'. The first deviation consists of a maximal  survey covering the 20,000 square degrees to a depth achieved by the SNAP deep survey (260 gal/arcmin$^2$).  This is a truly extreme case, since this geometry requires three and half thousand years of observation time with the SNAP satellite.  We also investigate a `shallow' survey which could be achieved within 3 years and would cover 20,000 square degrees to a depth of $\rm z_m=0.9$, which is close to DUNE and LSST. Table \ref{tbl:benchmarks} summarises the properties of these benchmark surveys. Figure \ref{fig:ideal} contains our results showing the 1 sigma errors ellipses on the $w_n - w_a$ plane by marginalizing over the other 5 parameters. 

\begin{table}
  \centering 
\begin{tabular}{c|c|c|c|c}
\hline
 Survey  & Area & $n_g$ & $z_m$ &FOM \\
   & sq. deg.  & gals/amin$^2$ &&\\

\hline
Shallow  & 20 000   & 35 &0.9&50 \\
Ideal   & 20 000   & 100 &1.23&220 \\
Maximal  & 20 000 & 260 &1.43&600\\
\hline
\end{tabular}
  \caption{Properties of the benchmark surveys.  Also shown is the figure of merit for each of the configurations.}
  \label{tbl:benchmarks}
\end{table}

\section{Impact of Limitations}

Having appreciated the potentials of an ideal lensing survey, we examine how external limitations and systematic errors diminish the accuracy of future lensing surveys. 

\subsection{Finite Observing Time}

The first limitation is that any future survey will be limited by a finite observing time. Within a realistic timeframe, of roughly 3 years, the ideal survey will be degraded in either depth or survey area.  Optimizing such a survey, therefore, needs a trade-off between these two competing parameters. There are three quantities that are important when considering survey geometry. These are the area of the survey, $A_s$; the number density of lensed galaxies, $n_g$; and the redshift distribution of the galaxies, which is well characterized by the median redshift of the distribution, $z_m$. 

\begin{figure} 
\centering
\resizebox{0.95\columnwidth}{0.8\columnwidth}{\includegraphics{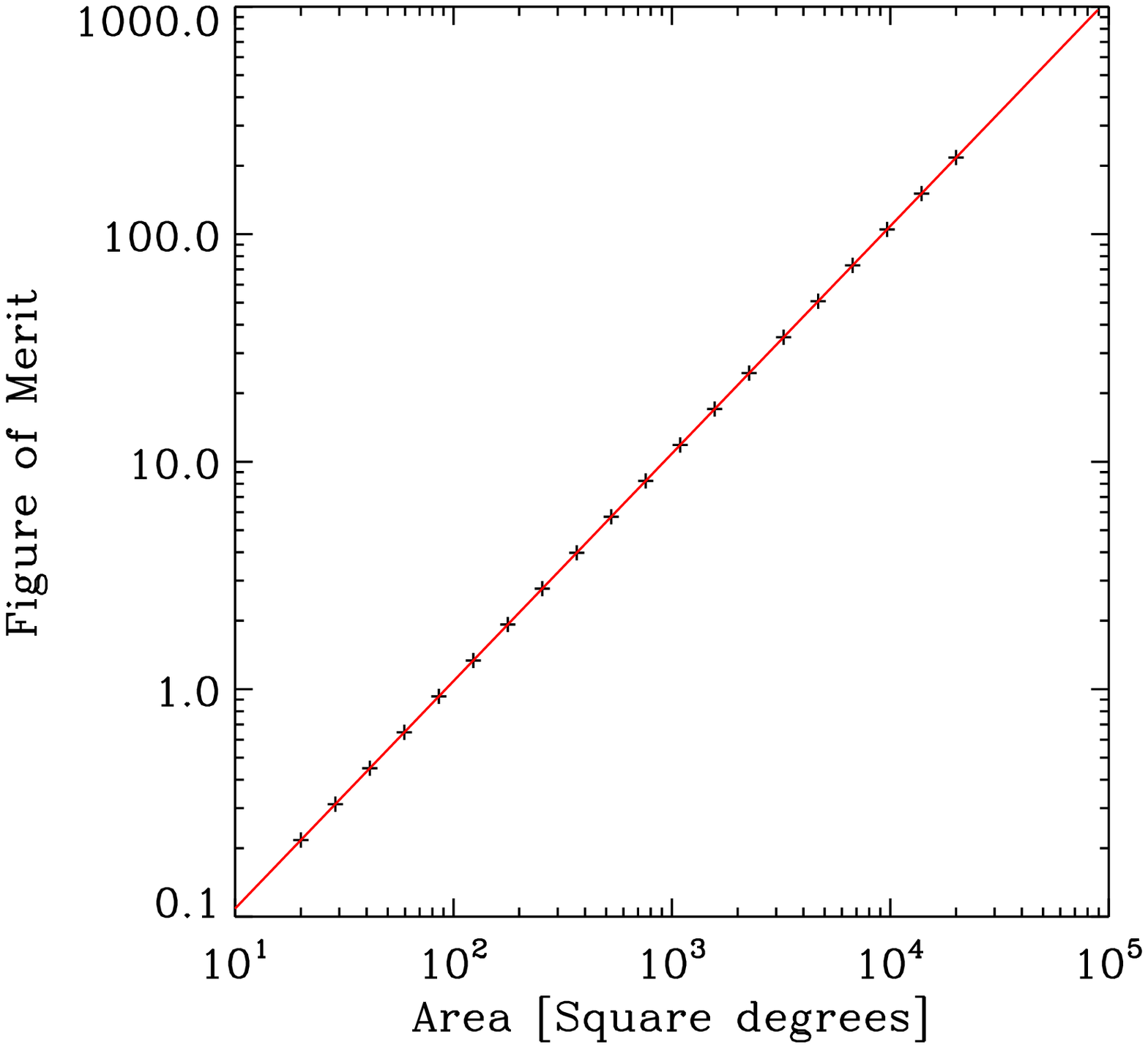}}
\resizebox{0.95\columnwidth}{0.8\columnwidth}{\includegraphics{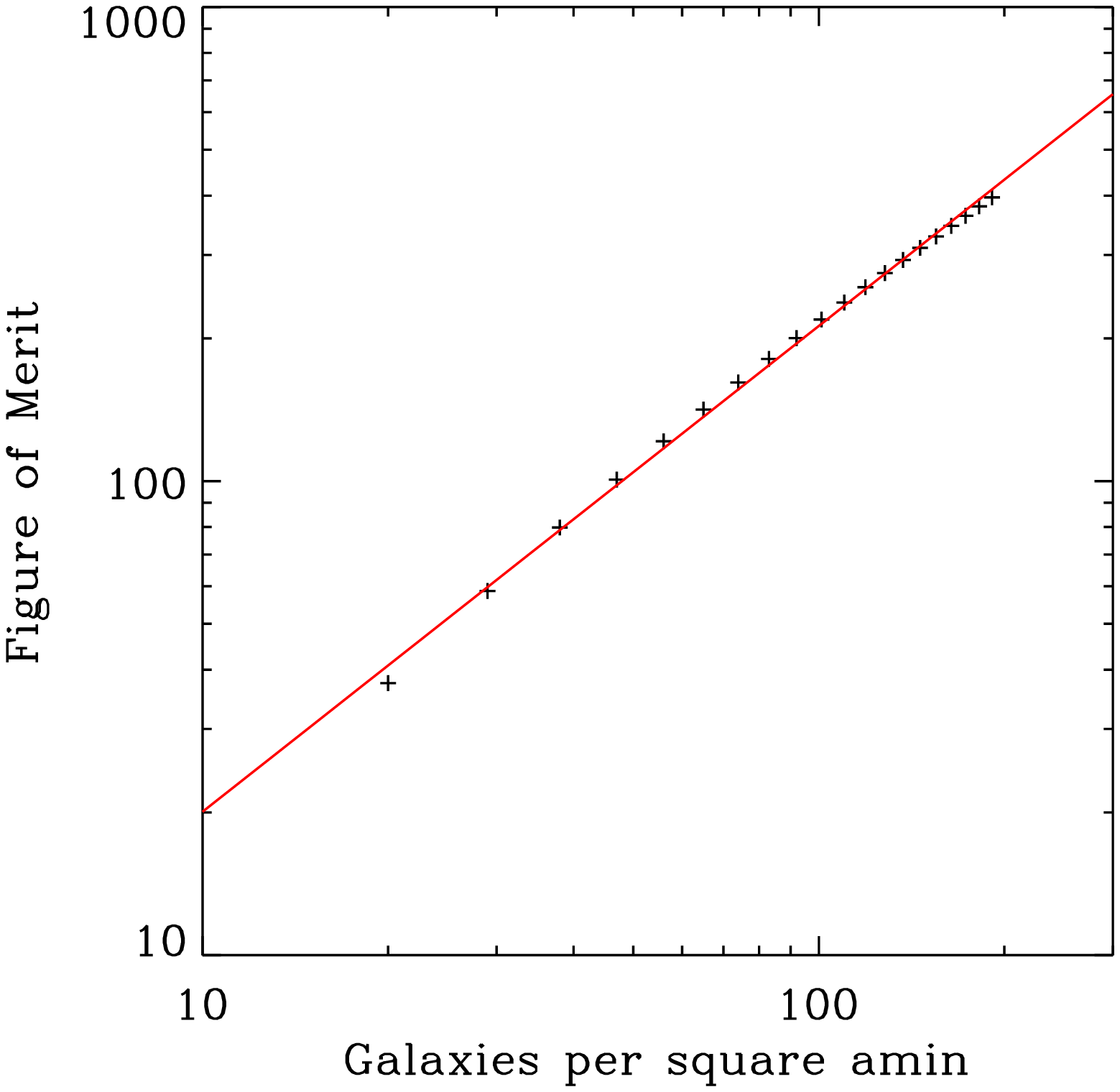}}
\resizebox{0.95\columnwidth}{0.8\columnwidth}{\includegraphics{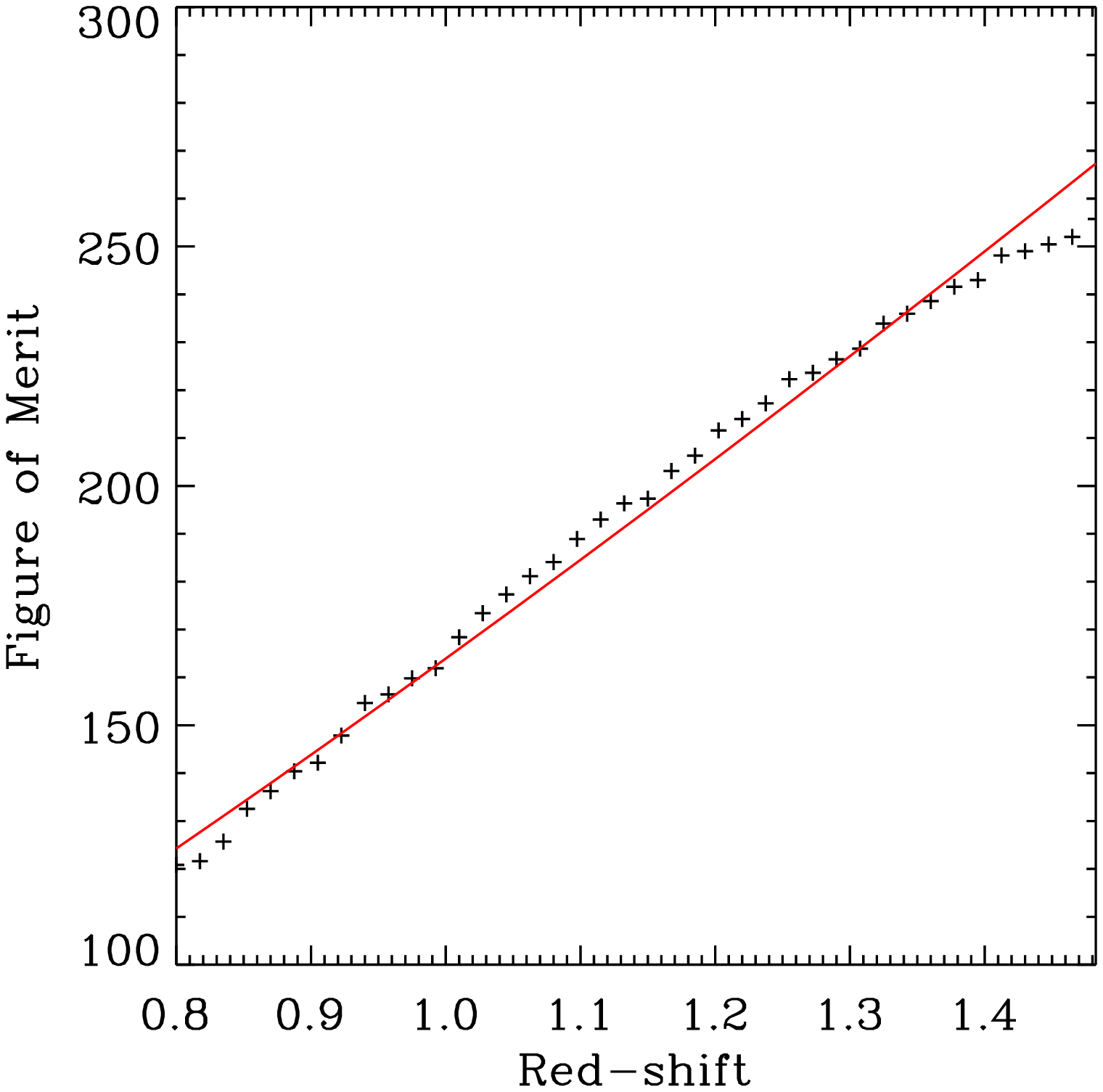}}
\caption{Effect of survey geometry on the DE FOM.  For each of the plots shown here only one survey property is varied at any time with others held fixed.  Top: Figure of merit as a function of survey area.  The symbols show the results of a Fisher calculation and the line shows a linear fit to the data. 
Middle: Figure of merit as a function of galaxy number density.  The redshift distribution is the same for all the points with a median redshift of 1.23.
Bottom: Dependence of the figure of merit on the median redshift of the lensed galaxies. The number density of galaxies has been fixed to 100 galaxies per square arc minute.}

\label{fig:fom_v_As} 
\end{figure}

\begin{figure} 
\centering
\resizebox{0.95\columnwidth}{0.95\columnwidth}{\includegraphics{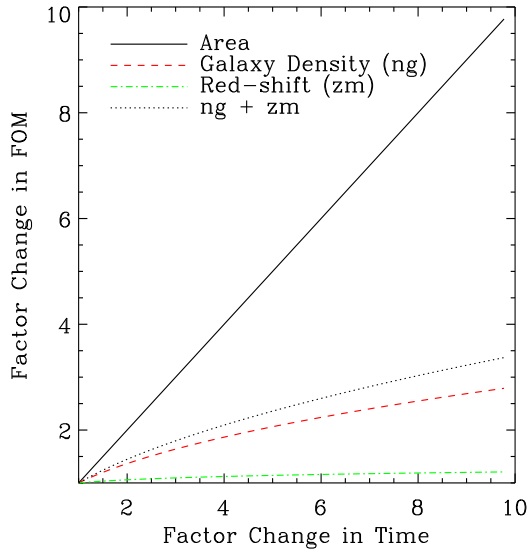}}
\caption{Gains in FOM when time is dedicated to increasing one of the three parameters which impact the statistics of cosmic shear.  We see that devoting observing time to increasing the area of the survey has the greatest impact on the FOM, while the change in median redshift lensed galaxies causes a minimal change in FOM.  When performing a deep vs. wide trade-off study, these three factors fall into two groups. Increasing the area requires observing time being spent going wide, while the other two factors prefer a deep survey. Taking this into account, we see that the gains from increasing area out-weigh the combined gains of zm and ng.}
\label{fig:dfom_dt} 
\end{figure}

Figure \ref{fig:fom_v_As} shows the impact of each of these parameters on the FOM.  Looking at each component separately, we find, by varying each parameter of the survey while fixing all the other parameters, that the figure of merit scales linearly with area ($FOM \propto A_s^{1.0}$). The scaling is also almost linear with galaxy density counts ($FOM \propto ng^{1.0}$) and has a stronger dependence on the median redshift ($FOM \propto z_m^{1.2}$).  These dependencies need to be cast in terms of observation time, which can only be fully evaluated using full image simulations for a survey.  These have been performed for both DUNE \citep{2006SPIE.6265E..58R} and SNAP \citep{2004AJ....127.3089M}. To estimate the dependency, we use the survey times quoted in table 1 of \cite{2004AJ....127.3102R}, which are the results of SNAP simulations.  We find that $z_m \propto t_{\rm obs}^{0.067}$ and $ng \propto t_{\rm obs}^{0.44}$.  Combining these effects, figure \ref{fig:dfom_dt} shows the fractional change in FOM as a function of the fraction of time dedicated to increasing $z_m$, $ng$ and $A_s$.  We see that the dominant property of a survey is its area.  Even when combining the benefits of a deep survey, i.e. higher $z_m$ and $ng$, we see that $A_s$ still dominates.

\begin{figure} 
\centering
\resizebox{0.95\columnwidth}{0.95\columnwidth}{\includegraphics{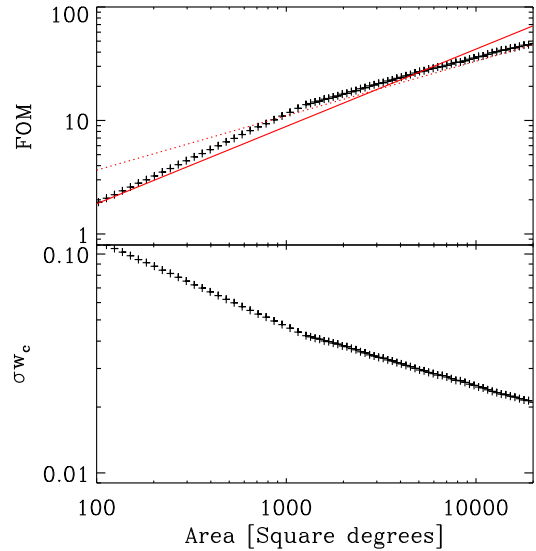}}
\caption{The results of a deep vs. wide trade-off study, given 3 years of observing time.  The survey area, galaxy number counts and their median redshift are calculated by interpolating  and extrapolating the results of \citet{2004AJ....127.3089M}. The three quantities are strongly correlated. Hence, a wide survey will have a lower galaxy number density and median redshift than a survey covering a small area.  The upper panel shows the optimisation using the FOM, quantifying the error levels on a 2 parameter  w model, and the lower panel shows the errors on the equation of state for a constant w model (i.e. a 1 parameter w model).  As discussed in section \ref{sec:theory_DEM}, improvement as measured by the FOM is greater than the improvement from a 1 parameter w model.}
\label{fig:deep_v_wide} 
\end{figure}

Finally, we perform a full deep vs wide trade-off study.  Figure  \ref{fig:deep_v_wide} shows the FOM for a survey limited to a mission time of three years with 60\% observing efficiency. Under these restrictions, the survey can either be wide, at the expense of depth or vice versa.  In order to correctly scale $A_s$, $n_g$ and $z_m$ we continue to use the space image simulation results of \cite{2004AJ....127.3089M}.  We have also confirmed that this scaling is consistent with the scaling found in the simulations described in \cite{2006SPIE.6265E..58R}. We see that once again a maximum FOM is achieved by the widest possible configuration, which is 20,000 square degrees.

\subsection{Photometric Redshifts}

For weak lensing tomography, the redshift distribution of the galaxies in each redhsift bin must be known. Spectroscopically measuring the redshifts of the galaxies over the entire area of future lensing surveys is not feasible. Hence, the next generation of lensing surveys will rely heavily on photometric redshifts, based on multi-color photometry.  The accuracy of this method depends on a number of factors including: 1) the accuracy of the photometry; 2) the number of bands; 3) the covered wavelength range; and 4) the number of spectroscopic redshift used for calibration.  These factors, along with the details of the specific mock or photo-z method used, impact the photometric errors in complicated ways.  Instead of tackling these details directly, we choose to model their impact in a more generic and simple way here. However, a more detailed analysis of specific imaging surveys, such as DES, Panstarrs and LSST, is in progress and will follow shortly (Abdalla et al in prep).

\begin{figure} 
\centering
\resizebox{0.95\columnwidth}{0.95\columnwidth}{\includegraphics{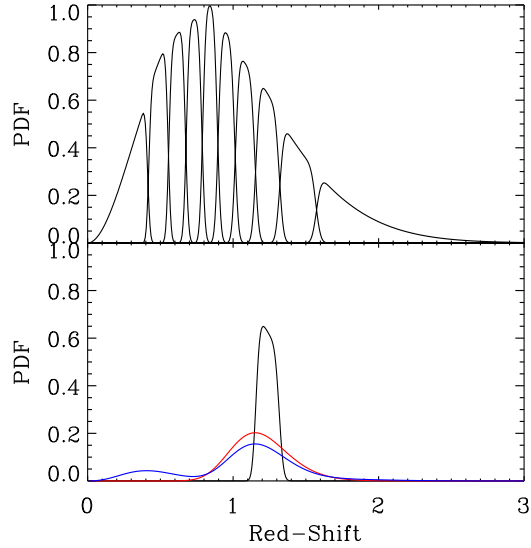}}
\caption{The upper panel shows the redshift distribution of galaxies, with $z_m=0.9$, divided into 10 redshift bins. The photometric redshifts are assumed to have $\delta_z=0.01$ and no catastrophic failures. The lower panel shows the distribution of the galaxies in the 8th redshift bin.  The red curve shows this redshift slice for $\delta_z=0.1$, and the blue curve also has $\delta_z=0.1$ and $f_{\rm cat \it}=0.3$.}
\label{fig:photoz1} 
\end{figure}

Here we consider 3 distinct effects: statistical dispersion, catastrophic failures and calibration errors for a given number of available spectroscopic redshifts.  We model the statistical dispersion in the measured redshift with a Gaussian that has a standard deviation of $\delta_s$. This is illustrated by the red curve in figure \ref{fig:photoz1}.  The catastrophic failures are modeled by adding extra Gaussians to the PDF of the galaxies, as shown by the the blue curve in figure \ref{fig:photoz1}, where the fraction of galaxies that suffer catastrophic failures is $f_{cat}$. Finally we use the number of galaxy spectra, $n_s$, to assess how well the mean and the variance of the redshift slices can be measured and hence calibrated.  The details of the model we use are described in detail in appendix \ref{ap1}.  

Figure \ref{fig:photoz1} shows the impact of these effects on the redshift slices. For our surveys, we assume that galaxies are distributed according to a PDF given by \cite{1994MNRAS.270..245S},
\begin{equation}
\label{ }
P(z)=z^\alpha \exp \bigg[-\bigg(\frac{z}{z_0}\bigg)^\beta\bigg],
\end{equation}
where we set $\alpha=2$ and $\beta=1.5$.  The median redshift of the survey, $z_m$, is then used to set $z_0 \simeq z_m/1.412$.

\begin{figure} 
\centering
\resizebox{0.95\columnwidth}{0.95\columnwidth}{\includegraphics{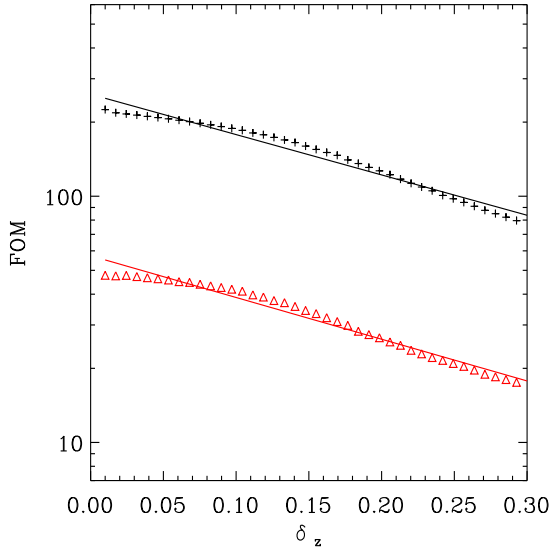}}
\caption{Impact of statistical errors in photometric redshift on the figure of merit parameter.  In black, we see the impact on the ideal survey, and in red we see the results for a low redshift survey ($z_m=0.9$).  The symbols show the results of the Fisher matrix analysis, and the lines show a linear fit to the data to highlight the trends. }
\label{fig:photoz2} 
\end{figure}

\begin{figure} 
\centering
\resizebox{0.95\columnwidth}{0.95\columnwidth}{\includegraphics{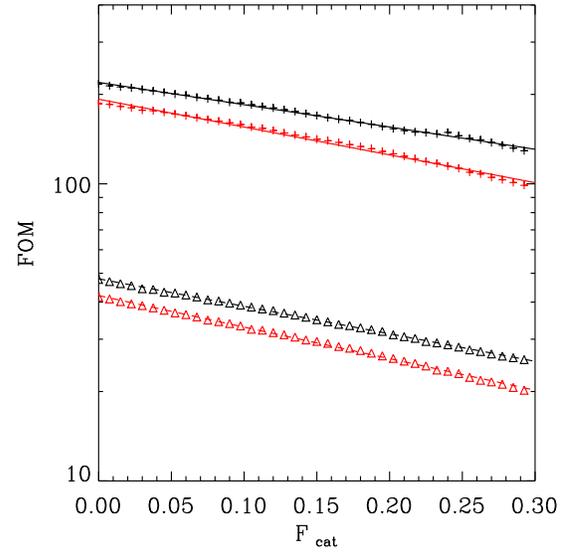}}
\caption{Impact of catastrophic failures of photometric redshifts on the figure of merit for four survey configurations. The results in black are for surveys with $\delta_z=0.01$, and the red results are for surveys with $\delta_z=0.1$.  The crosses show the Fisher matrix results of surveys that have the same geometry as our ideal survey.  The triangles show the results for a shallower survey, with $z_m=0.9$ and $ng=35$.  The lines show a linear fit.}
\label{fig:photoz_cat} 
\end{figure}

Figure \ref{fig:photoz2} shows the change in figure of merit as a function of $\delta_z$. We see a clear degradation of the FOM with increasing $\delta_z$.  We find that for our ideal survey, the figure of merit scales as, $FOM \propto 10 ^ {-1.64\delta_z}$.  We also find that for a shallower survey, with $z_m=0.9$ and $ng=35$, the figure of merit also shows a drop with $\delta_z$, $(FOM \propto 10^{-1.69\delta_z}$).  Investigating the impact of catastrophic failures, we also find a decrease in the figure of merit for an increase in the catastrophic failure fraction, $f_{\rm cat}$, (figure \ref{fig:photoz_cat}).  For the ideal survey, we find that $FOM \propto 10^{-0.75f_{\rm cat \it}}$.  We also find that for a survey with the same geometry as our ideal survey but with $\delta_z=0.1$, $FOM\propto 10^{-0.94f_{\rm cat \it}}$, the shallow survey with $\delta_z=0.01$ has a $FOM \propto 10^{-0.93f_{\rm cat \it}}$, and finally that a shallow survey with $\delta_z=0.1$ has $FOM\propto 10^{-1.1f_{\rm cat \it}}$.

\begin{figure} 
\centering
\resizebox{0.95\columnwidth}{0.95\columnwidth}{\includegraphics{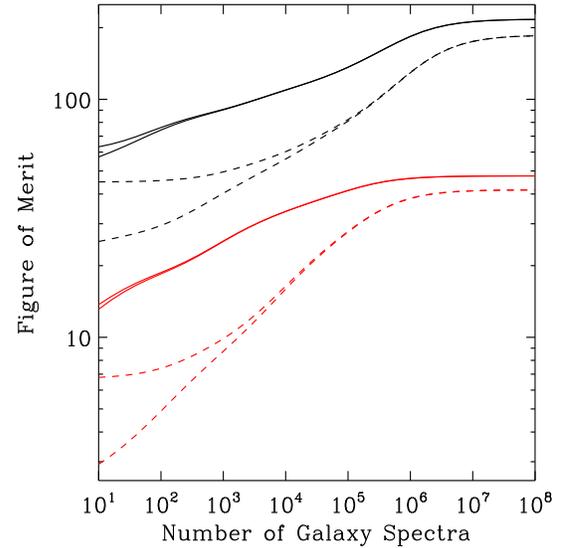}}
\caption{Importance of spectroscopic redshift measurements for cosmic shear tomography. Four cases are shown: (i) our ideal survey -black solid curves; (ii) our ideal survey with $\delta_z=0.1$ - black dashed curves; (iii) our shallow survey with $\delta_z=0.01$ - red solid curves; and (iv) our shallow survey with $\delta_z=0.1$ - red dashed curves.  For each case, two curves are shown.  The bottom curves correspond to marginalizing over both mean and variance of the distribution, while the top curves are the results when we marginalize only over the mean and fix the variance.}
\label{fig:photoz_spec} 
\end{figure}

Next we show the requirements on the calibration sample, namely how the FOM depends on the number of spectroscopic redshifts available, $n_s$.  Figure \ref{fig:photoz_spec} shows the results for four cases, (i) our ideal survey, (ii) our ideal survey with $\delta_z=0.1$, (iii) our shallow survey with $\delta_z=0.01$, and (iv) our shallow survey with $\delta_z=0.1$.  These values of $\delta_z$ have been chosen to look at the difference one would expect from a good photo-z survey ($\delta_z=0.01$) and a more modest survey ($\delta_z=0.1$). For each case we show the calculations for two scenarios: in the first we marginalize over both the mean and the variance, and in the second we marginalize only over the mean and fix the variance.  From our results we see that uncertainties in these quantities (mean and variance) are important and can substantially reduce the sensitivity of a survey, although it is interesting to note that even with a small number of calibration galaxies, weak lensing tomography is able to provide good self-calibration. For large number of galaxies in the range $10^{4} - 10^{5}$, the photometric calibration is robust, which is in agreement with \citet{2006ApJ...636...21M} who also find that they need this many galaxy spectra for calibration.  We see that in this region where very few galaxy spectra are available, uncertainties in both the mean and the variance play an important role.  However, if the number of calibration galaxies exceeds $10^4$, only the uncertainty in the mean is important.  This suggests that if we do investigate higher order moments, they should only be important when the number of galaxy spectra is small. We can expect that above $10^4$ galaxy spectra, the uncertainty in the mean will continue to dominate.

\subsection{Shear Measurement Systematics and Theoretical Uncertainties}

The final sources of error we consider are those associated with the lensing power-spectrum itself.  These errors could have a number of origins, ranging from residual galaxy shape correlations arising from imperfect PSF deconvolution to uncertainties in the theoretical predictions.  Due to the potential complexity and unknown nature of this error, we consider a simple phenomenological error model for the power spectrum. Specifically, we consider a systematic uncertainty in all the power-spectra (auto-correlations and cross-correlations) of the form,
\begin{equation}
\label{eq:error_sys}
\Delta C_{\ell, {\rm sys}}=\Delta C_{\ell_{o}}\Bigg(\frac{\ell}{\ell_{o}}\Bigg)^\beta,
\end{equation}
where $\Delta C_{\ell_{o}}$ controls the amplitude of the systematic error
at a reference multipole $\ell_{o}$.

We add this error to the statistical errors by creating a covariance matrix with the same dimensions as that of the statistical covariance matrix.  We then fill the diagonal elements of this systematic covariance matrix with the square of the error shown in equation \ref{eq:error_sys} and add this matrix to the statistical covariance matrix. This, therefore, assumes that these systematic errors are uncorrelated. 

\begin{figure} 
\centering
\resizebox{0.95\columnwidth}{0.95\columnwidth}{\includegraphics{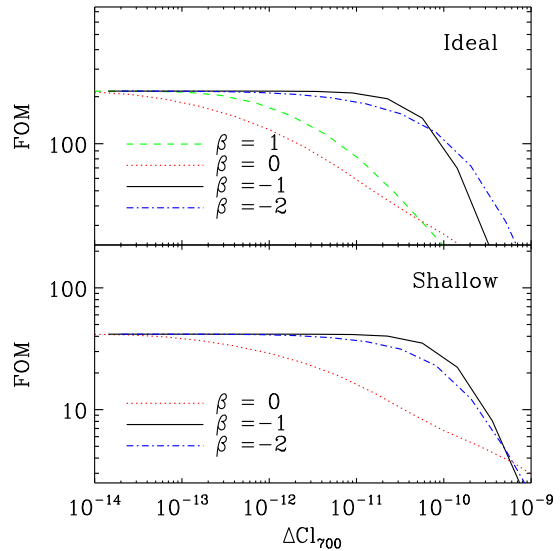}}
\caption{Impact of systematic errors in the lensing power-spectra on the FOM. The top panel shows the results for our ideal survey, while the lower panel shows those for the shallow survey (with ng=35, $z_m$=0.9 and $\delta_z$=0.1).  For each panel we show the FOM against $\Delta C_{\ell_{o}}$, the value of the systematic error at $\ell = 700$ for a number of slopes, $\beta$.  We see that for $\ell=700$, the impact of the error are largely in agreement up to $\sim~10^{-10}$.  This is not true for $\beta = -1$, which is the slope of the statistical error.  For this case the FOM is not affect by this form of error until they reach the same level as the statistical errors.}
\label{fig:sys1} 
\end{figure}

Figure \ref{fig:sys1} shows the results of adding these systematics to both the ideal survey and the shallow survey.  We find that, if we choose $\ell_{o}=700$, the impact of the systematic errors is largely independent of $\beta$ over a reasonably large range ($-2 <\beta < 1$).  This suggest that it is the level of the errors at $\ell\sim700$ that dominate.  The exception to this is $\beta=-1$, where the level of the error does not affect the figure of merit until $\Delta C_{\ell_{o}} \sim10^{-11}$ for the ideal survey and $\Delta C_{\ell_{o}} \sim10^{-10}$ for the shallow survey. This is because for $\beta=-1$ the systematic errors scale roughly the same way as the statistical errors (including cosmic variance) on these scale.  This means that the systematic errors are subdominant until they reach the same level as the statistical errors.  Beyond this point, the systematics errors dominate on all scales and, hence, have a large impact on the FOM.  We find that the scaling for the ideal and the shallow survey are well described by
\begin{equation}
\label{ }
FOM\propto \Delta C_{\ell_{o}}^{-0.16\big(\frac{\beta_{\dagger}^2}{\beta_{\dagger}^2+0.1}\big)}
\end{equation}
 and  
\begin{equation}
\label{ }
FOM\propto \Delta C_{\ell_{o}}^{-0.17\big(\frac{\beta_{\dagger}^2}{\beta_{\dagger}^2+0.1}\big)}, 
\end{equation}
respectively, where $\beta_{\dagger}=\beta+1$.  This is only valid in the range $2\times10^{-13} < C_{\ell_{o}} < 2\times10^{-10}$, with $\beta \neq -1$.  For $\beta=-1$, this is only valid in the ideal case for $\Delta C_{\ell_{o}} < 5\times10^{-11}$ and for the shallow survey for $\Delta C_{\ell_{o}} < 10^{-10}$.

\begin{figure} 
\centering
\resizebox{0.95\columnwidth}{0.95\columnwidth}{\includegraphics{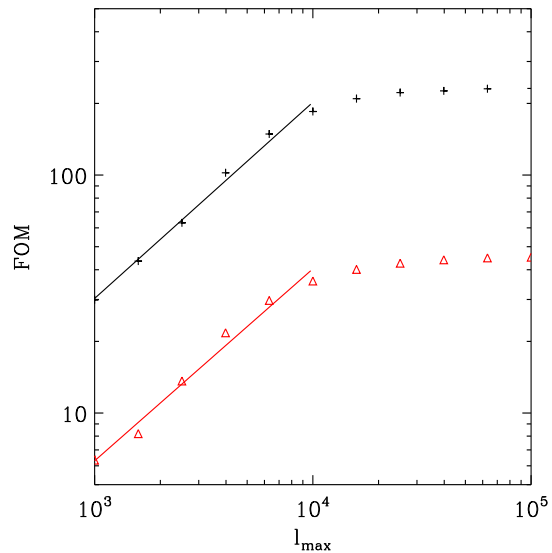}}
\caption{Figure of merit as a function of the maximum multipole $l_{\rm max}$ used for the
tomographic measurement. The minimum multipole is kept constant at $l_{\rm min}=10$.
The upper points (black crosses) are for the ideal survey, and the lower points (red triangles) are for a shallow survey with $z_m=0.9$, $n_g=35$ amin$^{-2}$ and $\delta_z=0.1$.}
\label{fig:sys2} 
\end{figure}

Another source of concern is the uncertainty of the theoretical predictions of the power spectrum on small scales.  In particular, the role played by baryons is not well understood at present \citep{2004ApJ...616L..75Z,2006ApJ...640L.119J}.  One simple way to circumvent this is to only fit within a scale range where the theoretical model is more secure. As explained in section \ref{sec:benchmark}, all the calculations desribed above assumed the range $10<\ell<2\times10^4$.  Figure \ref{fig:sys2} shows the FOM as a function of $\ell_{max}$, where the analysis has been performed in the range $10<\ell<\ell_{max}$.  We see the $\ell_{max}$ has a significant impact on the precision of the survey.  We find that for the ideal case $FOM \propto \ell_{max}^{0.27}$, but for a shallow survey with $\delta_z=0.1$ we find $FOM \propto \ell_{max}^{0.22}$. 

\section{Discussion}
In the preceding sections we have calculated the dependence of the FOM on a number of limiting parameters pertaining to survey properties and systematic effects. For this purpose, we considered deviations from set benchmark surveys, namely an ideal survey, which is both wide and deep, and a shallow survey, which is wide. The results are summarised by Equations \ref{eq:bigone} and \ref{eq:bigone2}, which give global analytic fits to the scaling of the FOM about the two respective surveys. These equations are convenient for the planning of future surveys, as it provides a first order optimization without needing to resort to the full Fisher matrix analysis. 

\begin{figure} 
\centering
\resizebox{0.95\columnwidth}{0.95\columnwidth}{\includegraphics{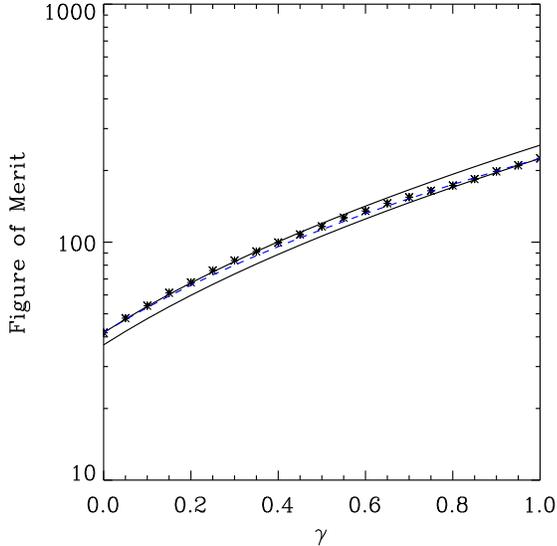}}
\caption{Comparison between the fitting functions shown in equations \ref{eq:bigone} (bottom black curve) and \ref{eq:bigone2} (top black curve). We use $\gamma$ to interpolate between the values of $FOM_1$ ($\gamma =1$) and $FOM_2$ ($\gamma=0$).  The points show the full Fisher matrix calculations.  The blue dashed curve shows the weighted average ($\gamma FOM_1+(1-\gamma) FOM_2$).  This figure also assumes that $n_s$ is sufficiently high in all cases so that the assumptotic values shown in figure \ref{fig:photoz_spec} are reached.}
\label{fig:fit_test} 
\end{figure}

We have also performed a test to verify that the two expressions are consistent with each other.  Figure \ref{fig:fit_test} shows the comparison between our two fitting functions (equation \ref{eq:bigone} and \ref{eq:bigone2}).  We use the parameter
$\gamma$ to interpolate $z_m$, $n_g$ and $\delta_z$ between the central models in equation \ref{eq:bigone} and equation \ref{eq:bigone2}, where $n_s$ has been set high enough that we reach the assymptotic values shown in figure \ref{fig:photoz_spec}. Noting that $\gamma =0$ corresponds to the values in equation \ref{eq:bigone2} and that $\gamma = 1$ corresponds to equation \ref{eq:bigone}, we see that, as expected, each fit work well close to point where it is supposed to. Also shown is a weighted average ($FOM_{\rm ave \it}=\gamma FOM_1+(1-\gamma) FOM_2$), which fits remarkably well over the entire range.

A further useful analysis is to look at the potential improvement in the FOM by considering a realistic range of survey parameters.  This is shown in Table \ref{tbl:ranges}.  For $A_s$, $n_g$ and $z_m$, the lower bounds are set by the CFHTLS \citep{2006ApJ...647..116H}, while the upper bounds are set by planned future missions, specifically, DUNE for the area and SNAP wide for the depth and number counts.  Next, we consider realistic ranges for the photometric redshift uncertainties. For this, we assume that current photometric redshifts have a statistical rms error of $10\%/(1+z)$ with  $10\%$ catastrophic failures and  can be calibrated using roughly 1000 spectroscopic measurements. To illustrate the room for improvement, we assume that future surveys will have a calibration set made up of $10^5$ spectroscopic galaxies, a statistical error of $1\%/(1+z)$ and no catastrophic failures.  This should be possible with a sufficient filter set that covers both the visible and the near infrared (NIR). 

For the $\ell$-mode cut off, we consider the range $10^3 < \ell_{max} <10^5$, which is the range that is typically considered in theoretical analyses and corresponds to discarding non-linear modes (lower bound) to including subarcmin scales (upper bound).  Finally, we consider the expected range of measurements and theoretical systematic errors. For the lower bounds in Table \ref{tbl:ranges} we consider recent measurements of $\sigma_8$.  These results show a scatter at the $10\%$ level, which is due to a combination of both statistical and systematic errors.  In these measurements, errors at typically quoted at the $5\%$ level ($1\sigma$) and include statistical errors and systematics, which would suggest that the remaining unaccounted for systematic errors are at the $8\%$ level. Given that there is currently $5\%$ uncertainty on the theoretical power-spectrum, we estimate that the total systematic errors on current $\sigma_8$ (or shear) measurements are at the $10\%$ level.  This, in turn, translates into a $20\%$ uncertainty in the shear power spectrum $C_\ell$.  Future lensing surveys have set $0.1\%$ as their target uncertainties on shear measurements \citep{2006MNRAS.366..101H,2006APh....26...91V,2006SPIE.6265E..58R}, $\gamma$, which translates into an uncertainty of $0.2\%$ in $C_\ell$. These systematic uncertainties can be converted into our $\Delta C_{\ell_o}$ through the values of $C_\ell$ at $\ell =700$, which we find to be $\sim 10^{-9}$ on average for the `ideal' configurations we have considered in this work.

We use equation \ref{eq:bigone} to estimate the impact on the FOM from the ranges shown in table \ref{tbl:ranges}.  We note, however, that the impact of the measurement and theory systematics part of the error budget is difficult to quantify accurately, since its extent will depend on the exact form of the uncertainty. To obtain the trends shown in equations \ref{eq:bigone} and \ref{eq:bigone2},  we have only considered uncorrelated errors, assuming a simple power-law form. We, therefore, consider this to be the lower bound of the impact. More work is needed to further refine these estimates.

From table \ref{tbl:ranges} we see that of three properties of a lensing survey, $A_s$, $n_g$ and $z_m$,  it is from the increased area that future surveys will most significantly improve over current measurements.  The impact of the other two properties falling far behind.  It is clear, therefore, that future surveys should first aim to increase area, then increase the number of galaxies and finally increase the median redshift of the lensed galaxies. From the table, we also see that the greatest gains in systematics can be achieved by increasing $\ell_{max}$, which is a threshold set by our uncertainties due to highly nonlinear interactions and, more importantly, baryonic physics.  We see that the other important systematic is the measurement/theoretical uncertainty in $C_\ell$, which will require: (1) improved numerical simulation of the nonlinear regime; (2) improvements in shear measurement techniques \citep{2006astro.ph..8643M}; and (3) tight control of image quality and PSF.  

Finally, we see that improvements in the photometric redshift errors will cause a small, but non-negligible, improvement in the FOM.  This improvement will come from two main fronts: (1) an increase in the number of galaxy spectra available for calibration; and (2) from reducing the spread of the errors, which can be achieved through better photometry measurements over a large filter set.  We see that relative to the other factors that have been considered, the effect of catastrophic failures is small. 

\begin{table*}
\begin{equation}
\label{eq:bigone}
FOM_{1}=140 \Bigg(\frac{A_s}{2\times10^4} \Bigg) \Bigg(\frac{n_g}{100} \Bigg)\Bigg(\frac{z_m}{1.23} \Bigg)^{1.2} \Bigg(\frac{n_s}{10^5} \Bigg)^{0.09} \Bigg(\frac{\Delta C_{\ell_{o}}}{2\times10^{-13} }\Bigg)^{{-0.16}\big(\frac{\beta_{\dagger}^2}{\beta_{\dagger}^2+0.1}\big)}  
\Bigg(\frac{\ell_{max}}{10^4} \Bigg)^{0.82} 10^{-1.6(\delta_z -0.01)} 10^{-0.75f_{\rm cat \it}}
\end{equation} 
\end{table*}
\suppressfloats

\begin{table*}
\begin{equation}
\label{eq:bigone2}
FOM_{2}=40 \Bigg(\frac{A_s}{2\times10^4} \Bigg) \Bigg(\frac{n_g}{35} \Bigg)\Bigg(\frac{z_m}{0.9} \Bigg)^{1.2} \Bigg(\frac{n_s}{10^5} \Bigg)^{0.24} \Bigg(\frac{\Delta C_{\ell_{o}}}{2\times10^{-13} }\Bigg)^{{-0.17}\big(\frac{\beta_{\dagger}^2}{\beta_{\dagger}^2+0.1}\big)}  
\Bigg(\frac{\ell_{max}}{10^4} \Bigg)^{0.81} 10^{-1.7(\delta_z -0.1)} 10^{-1.1f_{\rm cat \it}}
\end{equation}
\end{table*}

\begin{table}
  \centering 
\begin{tabular}{l|r|c|l|c|c}
\hline

 Survey & \multicolumn{3}{c}{Range}  &Fractional \\
 Property & & & & change in FOM \\
\hline
$A_s$ & $170$ &$\rightarrow$& $2\times 10^4$& 120\\
$n_g$ & $ 15$ &$\rightarrow$& $100 $ & 6.7\\
$z_m$ & $ 0.66$ &$\rightarrow$& $1.23 $  & 2.1\\
$n_s$ &  $ 10^3$ &$\rightarrow$&$ 10^5 $ & 1.5\\
$\delta_z$ &  $ 0.1$ &$\rightarrow$&$ 0.01$ & 1.4\\
$f_{\rm cat \it}$ &  $ 0.1$ &$\rightarrow$&$ 0.0 $ & 1.2\\
$\ell_{max}$ &  $ 10^3$ &$\rightarrow$&$ 10^4 $ & 6.6\\
$\Delta C_{\ell_{o}}/10^{-9}$ & $  0.2 $ &$\rightarrow$&$ 0.002$ & $\geq2.2$ \\
\hline
\end{tabular}
  \caption{ Table showing the fractional improvement in the FOM for a feasible range of survey properties using equation \ref{eq:bigone}.}
  \label{tbl:ranges}
\end{table}

\section{Conclusion}
In this paper, we have studied how future weak lensing surveys can be optimised to maximise the constraints on dark energy parameters using the figure of merit proposed by the Dark Energy Task Force. For this purpose, we have considered the impact on power spectrum tomography of the survey geometry and of various systematics, such as photometric redshift errors, shape measurement errors, and theoretical uncertainties. This was done by considering deviations from benchmark surveys to study the impact of each of these effects.

We found that, for a given observing time, it is always advantageous to choose a wide
rather than deep survey geometry. We also find that a wide shallow survey captures most of the dark energy information by reaching a FOM of about 110, which is only a factor of 5 less than an unrealistic maximal survey that is both wide and very deep.

We also found that the dark energy constraints from power spectrum tomography
are fairly robust to systematics. In particular, we extend and simplify the formalism used by \cite{2006ApJ...636...21M} for modeling the impact of photometric redshift errors.  Using this, we find that as long as a subsample of $10^4-10^5$ spectra are available for calibration, weak lensing tomography is robust to statistical errors and catastrophic failures in the photometric redshifts. The introduction of statistical dispersion, bias or catastrophic failures of photometric redshifts indeed only affect the
dark energy figure of merit at the $40\%$ level, which is consistent with the recent findings of \cite{2007JCAP...03...13J}. This is small compared to the limitations that come from uncertainties in the power spectrum, due to measurement-theoretical systematics or requiring a high $\ell$ cutoff, which, respectively, have a factor of 2.2 and 3.5 impact on FOM.

In our optimisation study, we have only considered power spectrum tomography.
Other statistics have been proposed which can improve upon the constraints
of power spectrum tomography. In particular, cross-correlation cosmography \citep{2004ApJ...600...17B} can separate the geometrical effects of dark energy from its effect on the growth of structure. Higher order statistics such as the bispectrum or peak statistics are also known to break degeneracies present when 2-point functions are used. While \cite{2006astro.ph..6416T} have shown that cross-correlation cosmography, at least when measured using detected haloes, also favors wide rather than deep surveys, it would be interesting  to study whether the optimisation results obtained here can be generalised to these more advanced statistics.  We have also assumed that our cosmic variance errors are Gaussian, which does not hold on small scales ($10^{3} < \ell <10^{5}$).  Including the effects of non-Gaussian errors is being investigated by other authors, and current results indicate that for the wide survey geometry studied here the impact of non-Gaussian errors is small (Takada et al - private communications). This, however, needs to be quantified further.  

\appendix
\section{Photometric Redshift Model}
\label{ap1}
The first uncertainty we consider is a statistical errors in measuring the photo-z.  We therefore begin by assuming that the redshift measured using photometry is related to the true redshift, $z_t$, through a probability distribution function, $P_{\rm stat \it}(z_t)$, that is Gaussian and centered on $z_t$.  This models the spread caused by photometric errors in the measurements. We are able to vary the level of the error through the standard deviation of $P_{\rm stat \it}$, which we assume scales as (1+$z$). This leads to:

\begin{equation}
\label{ }
\langle P_{\rm stat \it} \rangle = \int dz ~ z P_{\rm stat \it}(z) =z_t,
\end{equation}

\begin{equation}
\label{ }
\sigma^2(P_{\rm stat \it}) = \int dz ~ z^2 P_{\rm stat \it}(z) =\delta_z^2 (1 + z_t)^2,
\end{equation}
where the level of this error is $\delta_z$, which is dominated by the quality of the photometric measurements and the number of filter bands used, and $z_t$ is the true redshift of the galaxy.  This spread is illustrated by the red curve in figure \ref{fig:photoz1}. 

Another source of error is introduced if the wavelength range covered is insufficient to correctly identify features in the galaxy spectrum, such as the Balmer break. This misidentification leads to galaxies being assigned a redshift that is off-set by $\Delta_z$ from the the true redshift.  We thus construct a probability distribution function to describe these catastrophic failures, $P_{\rm cat \it}(z_t)$.  Since galaxies can either be misidentified with a redshift offset $\Delta_z$, that is either higher or lower than the true redshift, we can construct a generic form of $P_{\rm cat \it}(z_t)$ to be bimodal, $P_{\rm cat \it}(z_t)=P_{\rm cat \it}^+(z_t) + P_{\rm cat \it}^-(z_t)$, with $P_{\rm cat \it}^+(z_t)$ capturing the PDF information for the catastrophic failures at redshifts greater than $z_t$ and $P_{\rm cat \it}^-(z_t)$ capturing the information about the scatter to redshifts below $z_t$. Then,

\begin{equation}
\label{ }
\langle P_{\rm cat \it}^\pm \rangle =z_{\rm cat \it}^\pm=z_t\pm\Delta_z.
\end{equation}

The value $\Delta_z$ depends on the set of filters used and the spectral properties of the galaxies in the survey.  For simplicity we assume that $\Delta_z=1$ and that the uncertainty of catastrophic failure population is the same as that for a galaxy whose true redshift is $z_{\rm cat \it}$,
\begin{equation}
\label{ }
\sigma(P_{\rm cat \it}^\pm)=\delta_z (1 + z_{\rm cat \it}^\pm).
\end{equation}
Combining these two effects we can create the probability distribution function of photometric redshift given a galaxy with a true redshift $z_t$,
\begin{equation}
\label{ }
P(z_{\rm phot \it}|z_t) =  (1-f_{\rm cat \it}) P_{\rm stat \it}(z_t)+ f_{\rm cat \it} P_{\rm cat \it}(z_t),
\end{equation}
where $f_{\rm cat \it}$ describes the fraction of galaxies that are misidentified.  An example of this is the blue curve of figure \ref{fig:photoz1}. Using Bayes' theorem, this conditional probability can be converted into a full 2D PDF,
\begin{equation}
\label{ }
P(z_t,z_{\rm phot \it}) = P(z_{\rm phot \it}|z_t)  P(z_t),
\end{equation}
where $P(z_t)$ is the probability distribution of true redshifts of the galaxies in the survey. This can be used to calculate the underlying distribution of galaxies after a cut is applied to the measured photometric redshifts.  Cuts can be made by applying a window function, $W_i(z_{\rm phot \it})$, and obtain
\begin{equation}
\label{ }
P_i(z_t)=\int dz_{\rm phot \it} W_i(z_{\rm phot \it}) P(z_{\rm phot \it},z_t).
\end{equation}
This is very similar to the approach taken by \cite{2006ApJ...636...21M}.  

Thus far we have examined the impact of changing the shape of the PDFs of the galaxies in each redshift bin.  This, however, does not take into account the errors introduced from the uncertainties in the shapes of the PDFs.  This third uncertainty arises because the algorithms used to measure the photometric redshifts of galaxies are developed and tested using a finite subsample of galaxies whose redshifts are measured spectroscopically.  The first step in performing this calibration is to ensure that the sample of spectroscopically measured galaxies are a fair representation of the galaxies in the lensing survey, i.e. subject to the same selection criteria.  With this done, the accuracy with which we are able to measure the PDF of the galaxies per redshift bin depends on the number of galaxies in that redshift bin.  The number of spectroscopic galaxy redshifts that are needed has also been studied by \cite{2006ApJ...636...21M} by assuming a 64 parameter binned bias model.  

Here, we take a slightly different approach.  We first assume that the photo-z's are used to divide the distribution of galaxies into redshift bins (as shown in figure \ref{fig:photoz1}).  Since the bias can take any form and need not be monotonic, we do not try to constrain this directly in the photo-z code.  Instead we use the fact that the PDF in each bin will be sampled with a finite number of galaxies, and it is that which will determine the accuracy of the distribution in each bin. In principle, this approach can be used to place error bars on all the moments of the distribution, which can then be marginalized over, but for now we focus on the mean and the variance of the distribution.  

For each bin we introduce the mean and the variance of each as a free parameter, which increases the number of parameters in our Fisher matrix from 7 to 27.  We then add a prior to the Fisher matrix that depends on the number of spectroscopic galaxies in each redshift bin. The priors on the mean are calculated using
\begin{equation}
\label{ }
\sigma^2(\bar{z})= \sigma^2(z)/n_s ,
\end{equation}
where $\sigma^2(\bar{z})$ is the variance of the mean, $n_s$ is the number of spectrascopic galaxies in each bin. Similarly, the priors on the sample variance are calculated by
\begin{equation}
\label{ }
\sigma^2(\sigma^2(z))= \frac{\mu_4 -\sigma^4}{n_s},
\end{equation}
where $\mu_4$ is the fourth order moment of the galaxy distribution in each bin.

\section*{Acknowledgements}
The authors are grateful to  Wayne Hu and Dipak Munshi for useful discussions and Damien Le Borgne for sharing his photo-z code.

\bibliographystyle{astron}
\bibliography{../../mybib}
\end{document}